\documentclass[aps,showpacs,preprintnumbers,amsmath,amssymb,twocolumn,superscriptaddress,prx,floatfix,nofootinbib]{revtex4-2}

\usepackage{amsfonts}
\usepackage{amsmath}
\usepackage{amssymb}
\usepackage{graphicx}
\usepackage{dcolumn}
\usepackage{bm,bbm}
\usepackage{xkcdcolors}
\usepackage{tabularx}
\usepackage{epstopdf}
\usepackage{xcolor}
\usepackage{caption}
\DeclareCaptionJustification{justified}
{\justifying}
\usepackage{mathrsfs}
\usepackage{subfigure}
\usepackage{mathtools}
\usepackage{centernot}
\usepackage{float}
\usepackage[linesnumbered,lined,boxed,commentsnumbered,ruled,vlined]{algorithm2e}
\usepackage{algpseudocode}
\usepackage{verbatim}
\usepackage{comment}
\usepackage{ragged2e}
\usepackage{physics}
\usepackage{hyperref}
\usepackage{cleveref}
\usepackage{tikz}
\usepackage{multirow}
\usepackage{colortbl}
\usepackage[bottom]{footmisc}
\hypersetup{colorlinks=true, citecolor=orange, urlcolor=blue, linkcolor=magenta}

\newcommand{\qed}{\hfill \ensuremath{\Box}}

\definecolor{posedges}{RGB}{51,51,153}
\definecolor{negedges}{RGB}{204,79,92}

\begin{document}

\title{Assessing frustration in real-world signed networks: a statistical theory of balance}

\author{Anna Gallo}
\email{anna.gallo@imtlucca.it}
\affiliation{IMT School for Advanced Studies, Piazza San Francesco 19, 55100 Lucca (Italy)}
\affiliation{INdAM-GNAMPA Istituto Nazionale di Alta Matematica `Francesco Severi', P.le Aldo Moro 5, 00185 Rome (Italy)}
\author{Diego Garlaschelli}
\affiliation{IMT School for Advanced Studies, Piazza San Francesco 19, 55100 Lucca (Italy)}
\affiliation{INdAM-GNAMPA Istituto Nazionale di Alta Matematica `Francesco Severi', P.le Aldo Moro 5, 00185 Rome (Italy)}
\affiliation{Lorentz Institute for Theoretical Physics, University of Leiden, Niels Bohrweg 2, 2333 CA Leiden (The Netherlands)}
\author{Tiziano Squartini}
\affiliation{IMT School for Advanced Studies, Piazza San Francesco 19, 55100 Lucca (Italy)}
\affiliation{Scuola Normale Superiore, Piazza dei Cavalieri 7, 56126 Pisa (Italy)}
\affiliation{INdAM-GNAMPA Istituto Nazionale di Alta Matematica `Francesco Severi', P.le Aldo Moro 5, 00185 Rome (Italy)}

\date{\today}

\begin{abstract}
According to the so-called strong version of structural balance theory, actors in signed social networks avoid establishing triads with an odd number of negative links. Generalising, the weak version of balance theory allows for nodes to be partitioned into any number of blocks with positive internal links, mutually connected by negative links. If this prescription is interpreted rigidly, i.e. without allowing for statistical noise in the observed link signs, then most real graphs will appear to require a larger number of blocks than the actual one, or even to violate both versions of the theory. This might lead to conclusions invoking even more relaxed notions of balance. Here, after rephrasing structural balance theory in statistically testable terms, we propose an inference scheme to unambiguously assess whether a real-world, signed graph is balanced. We find that the proposed statistical balance theory leads to interpretations that are quite different from those derived from the current, deterministic versions of the theory.
\end{abstract}

\maketitle

\paragraph*{\bf Introduction.} The interest towards signed networks dates back to \emph{balance theory}, first proposed by Heider as a theory of behaviour~\cite{heider1946attitudes}. As the name suggests, the theory revolves around the concept of `balance', i.e. the tendency of people to engage, and remain, in situations that support their beliefs. The practice of adopting signed graphs to model social networks subsequently led Cartwright and Harary~\cite{cartwright1956structural} to introduce the structural version of balance theory~\cite{cartwright1956structural,harary2002signed,ou2015detecting,iorio2016efficient,saiz2017evidence}: a complete, signed graph is said to be balanced if \emph{all triads} have an even number of negative edges, i.e. either zero (in this case, the three edges are all positive) or two. The so-called \emph{structure theorem} states that a complete, signed graph is balanced if and only if its set of nodes can be partitioned into $k=2$, disjoint subsets whose intra-modular links are all positive and whose inter-modular links are all negative. Cartwright and Harary extended the definition of balance to incomplete graphs~\cite{cartwright1956structural} by including cycles of length larger than three: a network is, now, said to be balanced if \emph{all cycles} have an even number of negative edges (although the points of each subset are no longer required to be connected). Taken together, the criteria above define the so-called \emph{strong balance theory}.

In an attempt to make such a framework more applicable, Davis introduced the concept of \emph{$k$-balanced} networks: according to it, signed graphs are balanced if their set of nodes can be partitioned into $k\geq2$, disjoint subsets with positive, intra-modular links and negative, inter-modular links~\cite{davis1967clustering}. This generalised definition of balance has led to the formulation of the \emph{weak balance theory}, according to which triads whose edges are all negative are balanced as well, since each node can be thought of as a group on its own. From a mesoscopic perspective, however, both versions of the balance theory require the presence of positive blocks along the main diagonal of the adjacency matrix ($k=2$, according to the strong variant; $k>2$, according to the weak variant~\cite{doreian1996partitioning,anchuri2012communities}) and of negative, off-diagonal blocks. Taken together, the strong and the weak variants of the balance theory define what may be called \emph{traditional balance theory}: hence, $k$-balanced networks are traditionally balanced.

Currently, when real-world networks are used to test traditional balance theory, the possible presence of statistical noise in the observed link signs is not taken into account. As a result, a network produced by a $k$-balanced process might appear as requiring a larger number $k'>k$ of blocks to be consistent with the theory, hence favouring the weak over the strong version of the theory. Even more dramatically, there might not be any partition into blocks with the `ideal' sign assignments compatible with  traditional balance theory. The theory might, then, be erroneously dismissed in favour of looser alternatives such as the so-called \emph{relaxed balance theory}~\cite{doreian2009partitioning}, which allows for the blocks of the matrix to be connected with the `wrong' signs - raising, however, the problem that a block structure with arbitrary signs can always be found on any signed graph, thus not being truly informative about the tendency towards balance of real-world networks. Here, we re-cast the idea of balance theory within a statistical framework, thus allowing for noise in the empirical link signs, while attempting at identifying the underlying `denoised' signed block structure from which more robust conclusions can be drawn about the observed level of balance. As expected, we find that the proposed, statistical variant of the theory leads to conclusions that are quite different from those derived from the current, deterministic variants.\\

\paragraph*{\bf Setting up the formalism.} Each edge of a signed graph can be \emph{positive}, \emph{negative} or \emph{missing}: as we will focus on binary, undirected, signed networks, a generic entry of the signed adjacency matrix $\mathbf{A}$ will be assumed to read $a_{ij}=-1,0,+1$, with $a_{ij}=a_{ji}$, $\forall\:i<j$. To ease mathematical manipulations, let us employ Iverson's brackets (a notation ensuring all quantities of interest to be non-negative - see Appendix \hyperlink{AppA}{A}) and define the quantities $a_{ij}^-=[a_{ij}=-1]$, $a_{ij}^0=[a_{ij}=0]$, $a_{ij}^+=[a_{ij}=+1]$: the new variables are mutually exclusive, sum to 1 and induce the two matrices $\mathbf{A}^+$ and $\mathbf{A}^-$, satisfying $\mathbf{A}=\mathbf{A}^+-\mathbf{A}^-$ and $|\mathbf{A}|=\mathbf{A}^++\mathbf{A}^-$. The number of positive and negative links is defined as $L^+=\sum_{i=1}^N\sum_{j(>i)}a_{ij}^+$ and $L^-=\sum_{i=1}^N\sum_{j(>i)}a_{ij}^-$, respectively.\\

\paragraph*{\bf Traditional Balance Theory.} The top-down formulation of the traditional balance theory leads quite naturally to the definition of a score function for quantifying the `degree of compatibility' of a given partition with the theory itself. It is named \emph{frustration}\footnote{More formally, \emph{line index of imbalance}~\cite{harary1959measurement,traag2019partitioning}.} and reads

\begin{align}\label{eq:Findex}
F(\bm{\sigma})&=\sum_{i=1}^N\sum_{j(>i)}a_{ij}^-\delta_{\sigma_i,\sigma_j}+\sum_{i=1}^N\sum_{j(>i)}a_{ij}^+(1-\delta_{\sigma_i,\sigma_j})\nonumber\\
&=L_\bullet^-+L^+-L_\bullet^+\nonumber\\
&=L_\bullet^-+L_\circ^+
\end{align}
where $\bm{\sigma}\equiv\{\sigma_i\}$ stands for a vector of labels characterising a generic partition and $\delta_{\sigma_i,\sigma_j}$ is the Kronecker delta (i.e. $\delta_{\sigma_i,\sigma_j}=1$ if $\sigma_i=\sigma_j$ and $0$ otherwise). In words, $F(\bm{\sigma})$ counts the amount of misplaced links according to the traditional balance theory, i.e. the number of negative links within modules (indicated with a filled dot) plus the number of positive links between modules (indicated with an empty dot). The simplest, operative criterion for singling out a $k$-balanced partition is based upon the following theorem see~\cite{harary1953notion,davis1967clustering,zaslavsky1987balanced,anchuri2012communities} for similar results).\\

\noindent\textbf{Theorem I.} $F(\bm{\sigma})=0\Longleftrightarrow$ the partition $\bm{\sigma}$ is $k$-balanced.\\

\noindent \emph{Proof.} Sufficiency condition: $F(\bm{\sigma})=0\Longrightarrow$ the partition $\bm{\sigma}$ is $k$-balanced. Since $L_\circ^+\ge0$ and $L_\bullet^-\ge0$, $F(\bm{\sigma})=0$ implies that $L_\circ^+=0$ and $L_\bullet^-=0$; hence, the definition of $k$-balanced partition is satisfied. Necessity condition: the partition $\bm{\sigma}$ is $k$-balanced $\Longrightarrow F(\bm{\sigma})=0$. Since a $k$-balanced partition is defined by the presence of a clustering with $k$ subsets, no negative links within modules and no positive links between modules, $L_\bullet^-=0$ and $L_\circ^+=0$; hence, $F(\bm{\sigma})=0$. $\qed$\\

In words, the bare, numerical value $F(\bm{\sigma})$ can be thought of as acting in a threshold-like fashion, classifying the configurations characterised by $F(\bm{\sigma})=0$ as balanced and the configurations characterised by $F(\bm{\sigma})>0$ as frustrated. The criterion embodied by the $F$-test is, however, too strict for real-world networks, which are hardly (if ever) found to obey it: as table \ref{tabI} shows, in fact, none of the listed configurations satisfies it. As noticed in~\cite{doreian2009partitioning}, the block-structure defining the traditional balance theory is overly restrictive, dooming the vast majority of real-world, signed networks to be quickly dismissed as frustrated - in fact, observing one, misplaced link is enough to let one conclude that the theory is not obeyed.\\

\begin{table}[t!]
\centering
\begin{tabular}{l|c|c|c|c|c}
\hline
& $N$ & $L$ & $F(\bm{\sigma})$ & \multicolumn{2}{c}{$G(\bm{\sigma}|\alpha)$} \\
\hline
\hline
 & & & & $\alpha=0.2$ & $\alpha=0.8$ \\
\hline
\hline  
Fraternity~\cite{aref2019balance} & 16 & 40 & 1 & 0.2 & 0.4 \\
\hline
N.G.H. Tribes~\cite{aref2019balance} & 16 & 58 & 2 & 1.4 & 0.4 \\
\hline
Slovenian Parliament~\cite{doreian1996partitioning} & 10 & 45 & 2 & 0.4 & 0.8 \\
\hline
Monastery~\cite{aref2019balance} & 18 & 49 & 5 & 2.4 & 1.8 \\
\hline
Spanish School 2~\cite{ruiz2023triadic} & 182 & 866 & 69 & 43.4 & 42.4 \\
\hline
Spanish School 1~\cite{ruiz2023triadic} & 359 & 2048 & 153 & 44 & 61 \\
\hline
US Senate~\cite{aref2019balance} & 100 & 2461 & 247 & 166.8 & 56 \\
\hline
\hline
CoW, 1946-49~\cite{doreian2015structural} & 60 & 360 & 12 & 3.8 & 5.8 \\
\hline
CoW, 1950-53~\cite{doreian2015structural} & 72 & 437 & 11 & 5.6 & 5.4 \\
\hline
CoW, 1954-57~\cite{doreian2015structural} & 80 & 492 & 27 & 7 & 12.2 \\
\hline
CoW, 1958-61~\cite{doreian2015structural} & 101 & 613 & 25 & 6.4 & 14.6 \\
\hline
CoW, 1962-65~\cite{doreian2015structural} & 109 & 642 & 32 & 16.8 & 24.6 \\
\hline
CoW, 1966-69~\cite{doreian2015structural} & 111 & 607 & 24 & 11.8 & 15.6 \\
\hline
\hline
EGFR~\cite{aref2019balance} & 313 & 755 & 189 & 51.2 & 46.8 \\
\hline
Macrophage~\cite{aref2019balance} & 660 & 1897 & 316 & 91.4 & 77.2 \\
\hline
\hline
Bitcoin Alpha~\cite{aref2020multilevel} & 3775 & 14120 & 1399 & 337.9 & 585.6 \\
\hline
Bitcoin OTC~\cite{aref2020multilevel} & 5875 & 21489 & 3259 & 540.4 & 800.4 \\
\hline
\end{tabular}
\caption{Empirical amount of frustration, detected by searching for the partition minimising $F(\bm{\sigma})$, that characterises the listed, real-world networks: according to the $F$-test, none of them turns out to satisfy the traditional balance theory. The same result holds true even when employing the generalised definition of the frustration index (here, implemented by posing $\alpha=0.2$ and $\alpha=0.8$).}
\label{tabI}
\end{table}

\begin{figure*}[t!]
\centering
\includegraphics[width=\textwidth]{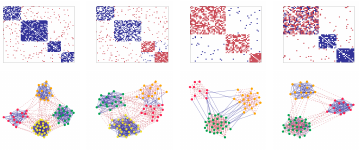}
\caption{Consistency checks on four, synthetic configurations (positive links are coloured in blue; negative links are coloured in red): minimising BIC always leads to recover the planted partition, be it homogeneous and balanced according to the traditional balance theory (first panel); homogeneous and balanced according to the relaxed balance theory (second and third panel); heterogeneous and balanced according to the statistical variant of the relaxed balance theory (fourth panel). Nodes of different colors belong to different groups.}
\label{fig1}
\end{figure*}

\paragraph*{\bf Softening frustration.} In order to overcome what was perceived as a major limitation of the traditional balance theory, Doreian and Mrvar~\cite{doreian2009partitioning} proposed to replace $F(\bm{\sigma})$ with its softened variant

\begin{equation}
G(\bm{\sigma}|\alpha)=\alpha L_\bullet^-+(1-\alpha) L_\circ^+,
\end{equation}
allowing \emph{i)} the misplaced, positive links to be weighted more upon choosing $0\leq\alpha<1/2$; \emph{ii)} the misplaced, negative links to be weighted more upon choosing $1/2<\alpha\leq1$. Even ignoring the ambiguity due to the lack of a principled way for selecting $\alpha$ (the so-called `$\alpha$ problem' in~\cite{traag2019partitioning}), the criterion embodied by the $G$-test is still too strict: as table \ref{tabI} shows, in fact, none of the listed configurations satisfies it either. This is rigorously stated by the following theorem, whose proof is immediate.\\

\noindent\textbf{Theorem II.} If $0<\alpha<1$, $F(\bm{\sigma})=0\Longleftrightarrow G(\bm{\sigma}|\alpha)=0$, i.e. the partition $\bm{\sigma}$ is $k$-balanced\footnote{Notice that the values $\alpha=0$ and $\alpha=1$ would, respectively, lead to the trivially balanced partition characterised by a single community gathering all nodes together and $N$ single-node communities (or singletons).}.\\

\paragraph*{\bf Relaxed Balance Theory.} In the light of the previous results, the second attempt pursued by Doreian and Mrvar~\cite{doreian2009partitioning} to overcome the perceived limitations of the traditional balance theory was more radical, as they proposed to relax it by allowing for the presence of positive, off-diagonal blocks and negative, diagonal blocks - a generalisation that has gained the name of \emph{relaxed balance theory}~\cite{doreian2009partitioning}. Such a formulation, however, lacks a proper mathematisation, as a score function such as $F(\bm{\sigma})$ or $G(\bm{\sigma}|\alpha)$ cannot be easily individuated. Besides, it is affected by the problem highlighted in~\cite{traag2009community}: `\emph{[\dots] if the number of clusters is left unspecified a priori, the best partition is the singletons partition (i.e. each node in its own cluster) [\dots]}'. A bit provocatively, one may say that `the remedy seems worse than the evil' as no criterion is provided to \emph{i)} quantify the extent of the violation of the traditional balance theory and \emph{ii)} assess if it is indeed so relevant to justify the introduction of an alternative, conceptual framework.\\

\paragraph*{\bf Statistical Balance Theory.} Re-casting the theory of balance within a statistical framework solves all the aforementioned problems at once, allowing us to define an inference scheme to unambiguously assess if a signed graph is either traditionally or relaxedly balanced - hence, overcoming the limitations of the $F$-based and $G$-based tests while providing a proper mathematisation of the relaxed balance theory.

In order to define a statistical theory of balance, let us suppose the presence of a probabilistic model behind the appearance of any signed configuration: the traditional balance theory could be, then, rephrased by posing 

\begin{align}
p_{rr}^-&=0,\quad r=1\dots k
\end{align}
and

\begin{align}
p_{rs}^+&=0,\quad\forall\:r<s,
\end{align}
with $p_{rr}^+$ indicating the probability that any two nodes belonging to the same block $r$ are connected by a positive link, $p_{rs}^+$ indicating the probability that any two nodes belonging to the different blocks $r$ and $s$ are connected by a positive link and analogously for their negative counterparts.

The starting point of our approach is that of softening these positions, replacing them with the milder relationships

\begin{align}
\text{sgn}[p_{rr}^+-p_{rr}^-]=+1,\quad r=1\dots k
\end{align}
which amounts at requiring $p_{rr}^+>p_{rr}^-$, $r=1\dots k$ and

\begin{align}
\text{sgn}[p_{rs}^+-p_{rs}^-]=-1,\quad\forall\:r<s
\end{align}
which amounts at requiring $p_{rs}^+<p_{rs}^-$, $\forall\:r<s$. A configuration satisfying these relationships will be claimed to support the statistical variant of the traditional balance theory: specifically, its strong variant if $k=2$ and its weak variant if $k>2$; otherwise (because $p_{rr}^+\leq p_{rr}^-$ for some, diagonal blocks or $p_{rs}^+\geq p_{rs}^-$ for some, off-diagonal blocks), it will be claimed to support the statistical variant of the relaxed balance theory. Additionally, we call a partition \emph{homogeneous} if either $p_{rs}^+=0$ or $p_{rs}^-=0$, $\forall\:r\leq s$; otherwise, it will be called \emph{heterogeneous} (see Appendix \hyperlink{AppB}{B}). In words, the deterministic rules firstly defined by Cartwright, Harary and Davis are replaced by a probabilistic criterion individuating `a tendency' to obey, or not to obey, the traditional balance theory.

Tuning the aforementioned parameters on a given, signed network requires a generative model to be specified: here, we will adopt the Signed Stochastic Block Model (considered also in~\cite{yang2015bayesian,jiang2015stochastic,yang2017stochastic} but derived within the Exponential Random Graphs framework in Appendix \hyperlink{AppB}{B}), defined by the likelihood function

\begin{align}
\mathcal{L}_\text{SSBM}=&\prod_{r=1}^k(p_{rr}^+)^{L_{rr}^+}(p_{rr}^-)^{L_{rr}^-}(1-p_{rr}^+-p_{rr}^-)^{\binom{N_r}{2}-L_{rr}}\nonumber\\
&\prod_{r=1}^k\prod_{s(>r)}(p_{rs}^+)^{L^+_{rs}}(p_{rs}^-)^{L^-_{rs}}(1-p_{rs}^+-p_{rs}^-)^{N_rN_s-L_{rs}},
\end{align}
where $N_r$ is the number of nodes constituting block $r$, $L_{rr}^+$ ($L_{rr}^-$) is the number of positive (negative) links within block $r$, $L_{rs}^+$ ($L_{rs}^-$) is the number of positive (negative) links between blocks $r$ and $s$, $\forall\:r<s$ and the probability coefficients read $p_{rr}^+=2L_{rr}^+/N_r(N_r-1)$, $p_{rr}^-=2L_{rr}^-/N_r(N_r-1)$, $r=1\dots k$ and $p_{rs}^+=L_{rs}^+/N_rN_s$, $p_{rs}^-=L_{rs}^-/N_rN_s$, $\forall\:r<s$. As maximising the bare likelihood is a recipe known to be affected by a number of limitations~\cite{traag2019partitioning}, we have opted for the minimisation of

\begin{align}
\text{BIC}&=\kappa_\text{SSBM}\ln n-2\ln\mathcal L_\text{SSBM},
\end{align}
named \emph{Bayesian Information Criterion}. Derivable as the saddle-point approximation of the (Bayesian) evidence of a model, such a criterion embodies a trade-off between parsimony (accounted for by the first addendum, with $\kappa_\text{SSBM}$ being the number of parameters of the model\footnote{To avoid confusion with the number of modules, $k$, characterising $k$-balanced networks, we have indicated the number of a model parameters as $\kappa$: naturally, $\kappa_\text{SSBM}=k(k+1)$ since we need to estimate two parameters per module.} and $n=N(N-1)/2$ proxying the system dimensions) and accuracy (accounted for by the second addendum, i.e. the log-likelihood term)~\cite{raftery1995bayesian,konishi2008information}. Although the magnitude of both addenda rises with the number of parameters, the log-likelihood term drives BIC towards more negative values while the parsimony term drives BIC towards more positive values: the number of parameters in correspondence of which the minimum is reached is selected and drives the network partition.

Our bottom-up approach is `maximally agnostic' towards any theory of balance, letting the data determine the number and the values of the parameters best fitting a given configuration: BIC is, in fact, sensitive to the `signed density' of the modules `by design', hence capable of spotting the presence of groups of nodes as well as attributing to each of them the sign of the majority of its constituting links. From a purely computational perspective, instead, the complexity of the algorithm to minimise BIC decreases with the link density $c=2L/N(N-1)$: in words, the denser the configuration, the faster the algorithm (see Appendix \hyperlink{AppB}{B}).\\

\begin{figure}[t!]
\centering
\includegraphics[width=0.49\textwidth]{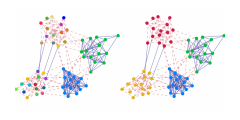}
\caption{Partitions recovered upon minimising $F(\bm{\sigma})$ (left panel) and upon minimising BIC (right panel): while minimising $F(\bm{\sigma})$ leads to recover a partition that is compatible with the traditional balance theory even if there is none `by design', minimising BIC leads to recover the homogeneous planted partition, compatible with the relaxed balance theory. Positive links are coloured in blue; negative links are coloured in red. Nodes of different colors belong to different groups.}
\label{fig2}
\end{figure}

\begin{figure*}[t!]
\centering
\includegraphics[width=\textwidth]{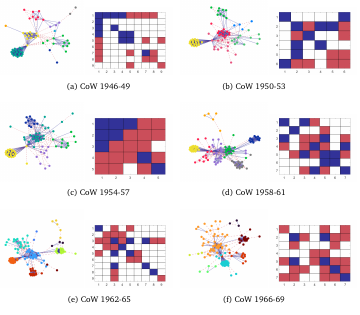}
\caption{Partitions recovered upon minimising BIC on six snapshots of the CoW dataset~\cite{doreian2015structural}, providing a picture of the international, political relationships over the years 1946-1997. A generic block, indexed as $rs$, is coloured in blue if $L_{rs}^+>L_{rs}^-$, in red if $L_{rs}^+<L_{rs}^-$ and in white if $L_{rs}^+=L_{rs}^-$: as blue blocks do not appear only on the diagonal and red blocks do not appear only off-diagonal, the considered configurations obey the statistical variant of the relaxed balance theory. Nodes of different colors belong to different groups.}
\label{fig3}
\end{figure*}

\paragraph*{\bf Results.} First, let us test our prescription on a number of synthetic configurations. As fig. \ref{fig1} shows, BIC minimisation always leads to recover the planted partition, irrespectively from the values of the sets of coefficients $\{p_{rr}^+\}$, $\{p_{rr}^-\}$, $\{p_{rs}^+\}_{r<s}$, $\{p_{rs}^-\}_{r<s}$, i.e. be it a homogeneous partition, balanced according to the weak variant of the traditional balance theory (more precisely, a 4-balanced partition); two, homogeneous partitions, balanced according to the relaxed balance theory (e.g. the third adjacency matrix is defined by $p_{11}^+=p_{22}^+=p_{33}^+=0$ and $p_{12}^-=p_{13}^-=p_{23}^-=0$); an heterogeneous partition, balanced according to the statistical variant of the relaxed balance theory (i.e. the fourth adjacency matrix, defined by $p_{11}^+<p_{11}^-$, $p_{22}^+>p_{22}^-$, $p_{33}^+>p_{33}^-$ and $p_{12}^+<p_{12}^-$, $p_{13}^+<p_{13}^-$, $p_{23}^+<p_{23}^-$). Additional exercises of the kind are reported in Appendix \hyperlink{AppB}{B}, where it is shown that employing BIC leads to robust inference towards perturbations aimed at degrading a given, balanced partition, across a wide range of the related parameters values.

Second, let us compare our recipe with that prescribing to minimise $F(\bm{\sigma})$. As fig. \ref{fig2} shows, implementing the latter does not lead to recover the planted partition (in this case, a homogeneous one, compatible with the relaxed balance theory): instead, it leads to a traditionally balanced configuration where the planted, negative cliques have been fragmented into singletons. Minimising $F(\bm{\sigma})$ can lead to a number of ambiguous situations, such as \emph{i)} returning configurations that are neither traditionally nor relaxedly balanced; \emph{ii)} returning more than one frustrated configuration (see Appendix \hyperlink{AppB}{B}). More in general, ignoring the interplay between the signs and the density of connections, solely accounting for the information carried by the first ones, may lead to `resolution errors' such as \emph{i)} splitting modules (even fully-connected ones) into finer regions or \emph{ii)} misinterpreting adjacent modules, characterised by the same, dominant sign, as single, coarser regions.

Let us, now, apply our recipe to a number of real-world, signed configurations, i.e. six snapshots of the `Correlates of Wars' (CoW) dataset~\cite{doreian2015structural}, providing a picture of the international, political relationships over the years 1946-1997 and consisting of 13 snapshots of 4 years each: a positive edge between any two countries indicates an alliance, a political agreement or the membership to the same governmental organisation; conversely, a negative edge indicates that the two countries are enemies, have a political disagreement or are part of different, governmental organisations. As fig. \ref{fig3} shows, minimising BIC leads to recover partitions that obey the statistical variant of the relaxed balance theory (a blue block is characterised by a majority of positive links; a red block is characterised by a majority of negative links); other real-world, signed configurations, instead, are found to obey the statistical variant of the traditional balance theory (see Appendix \hyperlink{AppB}{B}). All such partitions are heterogeneous. Overall, larger configurations seems to align more with the (statistical variant of the) relaxed balance theory while smaller configurations seems to align more with the (statistical variant of the) traditional balance theory.\\

\paragraph*{\bf Discussion.} The present paper proposes a statistical approach to the theory of balance, assuming that any real-world, signed configuration is the result of a generative process, probabilistic in nature. As some, unavoidable, degree of statistical noise is expected to affect any observed network structure, the criterion adopted to assess the consistency with balance theory can be re-cast in terms of the signs of the \emph{estimated probabilities} to observe positive and negative links, i.e. $p_{rs}^+-p_{rs}^-$, $\forall\:r\leq s$. Estimating these coefficients by minimising BIC allows one to unambiguously assess which variant of the theory is obeyed, from a statistical perspective, by any signed configuration.

On the contrary, minimising $F(\bm{\sigma})$ is practically equivalent to carrying out a sort of one-sided test of hypothesis, allowing one to  conclude if a given partition \emph{does not obey} the traditional balance theory (as a matter of fact, practically always) but incapable of providing a univocal classification for a generic, signed configuration. Moreover, it `works' even with configurations generated by the Signed Random Graph Model, i.e. a model carrying no information about a network modular structure, hence \emph{overfitting} (i.e. misinterpreting statistical noise as a genuine signal - see Appendix \hyperlink{AppB}{B}).

Under this respect, maximising the \emph{signed modularity} $Q(\bm{\sigma})$ is of no help, being defined as

\begin{align}
Q(\bm{\sigma})&=\sum_{i=1}^N\sum_{j(>i)}[(a_{ij}^+-p_{ij}^+)-(a_{ij}^--p_{ij}^-)]\delta_{\sigma_i,\sigma_j}\nonumber\\
&=-F(\bm{\sigma})+\langle F(\bm{\sigma})\rangle+L^+-\langle L^+\rangle
\end{align}
with obvious meaning of the symbols (the addendum $L^+-\langle L^+\rangle$ is just an offset not depending on the specific partition and amounting at zero for any model reproducing the total number of positive links)~\cite{gomez2009analysis}. In words, the signed modularity compares the empirical amount of frustration of a given, signed configuration with the one predicted by a properly-defined reference model: one may, thus, define a partition as \emph{statistically balanced} if satisfying the relationship $F(\bm{\sigma})<\langle F(\bm{\sigma})\rangle$, i.e. $Q(\bm{\sigma})>0$. Although reasonable, such a criterion does not differ (much) from the one embodied by the $F$-test: more formally, it can be proven that the relationship $L^+\gg L^-$ (often, if not always, found to hold true for real-world, signed networks) favours the fragmentation of the negative cliques into singletons, hence leading to recover traditionally balanced configurations even when there is none `by design' (see fig. \ref{fig2} and Appendix \hyperlink{AppC}{C}).\\

\paragraph*{\bf Conclusions.} Although the problem of partitioning a signed network has been approached in the past (e.g. by maximising the signed modularity~\cite{anchuri2012communities,gomez2009analysis,amelio2013community}), the existing works have completely overlooked the issue of harmonising the request of having balanced configurations with that of having modular configurations, in most of the cases verifying either the `degree of balance' of modular structures or the `degree of modularity' of balanced structures \emph{a posteriori}~\cite{esmailian2015community,su2017algorithm}; generative models, instead, can accommodate both requests, thus avoiding to return contradictory results - i.e. those one generally gets when combining a purely structure-based community detection with a purely sign-based one - while laying the basis of a more comprehensive theory of balance, grounded on probability theory (see~\cite{gallo2024testing,hao2024proper,gallo2024patterns} for related results).

Re-casting the idea of balance within a statistical framework allows the presence of noise in the empirical link signs to be accounted for, hence overcoming the limitations of the current, deterministic theories, which are doomed to misinterpret random patterns for genuine signals of (im)balance. Our inference scheme, instead, allows us to unambiguously assess whether a real-world, signed graph obeys the traditional notion of balance or aligns with a more relaxed variant of it. As a last point, we would like to stress that, within such a framework, the standard notion of frustration should be replaced by a fuzzy one, to be interpreted as proxying the `distance' of a given configuration from the closest, either traditionally or relaxedly, balanced one.\\

\textbf{Acknowledgments.} This work is supported by the European Union - NextGenerationEU - National Recovery and Resilience Plan (Piano Nazionale di Ripresa e Resilienza, PNRR), project `SoBigData.it - Strengthening the Italian RI for Social Mining and Big Data Analytics' - Grant IR0000013 (n. 3264, 28/12/2021). This work is also supported by the project `Reconstruction, Resilience and Recovery of Socio-Economic Networks' RECON-NET EP\_FAIR\_005 - PE0000013 `FAIR' - PNRR M4C2 Investment 1.3, financed by the European Union – NextGenerationEU. DG acknowledges support from the Dutch Econophysics Foundation (Stichting Econophysics, Leiden, the Netherlands) and the Netherlands Organization for Scientific Research (NWO/OCW).

\bibliography{bibmain.bib}

\clearpage

\onecolumngrid

\hypertarget{AppA}{}
\section*{Appendix A.\\Representing binary, undirected, signed networks}

The three functions $a_{ij}^-=[a_{ij}=-1]$, $a_{ij}^0=[a_{ij}=0]$ and $a_{ij}^+=[a_{ij}=+1]$ have been defined via the Iverson's brackets notation. Iverson's brackets work in a way that is reminiscent of the Heaviside step function, i.e. $\Theta[x]=[x>0]$; in fact,

\begin{equation}
a_{ij}^-=[a_{ij}=-1]=\begin{dcases}
1, & \text{if}\quad a_{ij}=-1\\
0, & \text{if}\quad a_{ij}=0,+1
\end{dcases}
\end{equation}
(i.e. $a_{ij}^-=1$ if $a_{ij}<0$ and zero otherwise),

\begin{equation}
a_{ij}^0=[a_{ij}=0]=\begin{dcases}
1, & \text{if}\quad a_{ij}=0\\
0, & \text{if}\quad a_{ij}=-1,+1
\end{dcases}
\end{equation}
(i.e. $a_{ij}^0=1$ if $a_{ij}=0$ and zero otherwise),

\begin{equation}
a_{ij}^+=[a_{ij}=+1]=\begin{dcases}
1, & \text{if}\quad a_{ij}=+1\\
0, & \text{if}\quad a_{ij}=-1,0
\end{dcases}
\end{equation}
(i.e. $a_{ij}^+=1$ if $a_{ij}>0$ and zero otherwise). These new variables are mutually exclusive, i.e. $\{a_{ij}^-,a_{ij}^0,a_{ij}^+\}=\{(1,0,0),(0,1,0),(0,0,1)\}$ and sum to 1, i.e. $a_{ij}^-+a_{ij}^0+a_{ij}^+=1$. The matrices $\mathbf{A}^+\equiv\{a_{ij}^+\}_{i,j=1}^N$ and $\mathbf{A}^-\equiv\{a_{ij}^-\}_{i,j=1}^N$ remain naturally defined, inducing the relationships $\mathbf{A}=\mathbf{A}^+-\mathbf{A}^-$, i.e. $a_{ij}=a_{ij}^+-a_{ij}^-$, $\forall\:i\neq j$ and $|\mathbf{A}|=\mathbf{A}^++\mathbf{A}^-$, i.e. $|a_{ij}|=a_{ij}^++a_{ij}^-$, $\forall\:i\neq j$.

\clearpage

\hypertarget{AppB}{}
\section*{Appendix B.\\Minimisation of the Bayesian Information Criterion}

\subsection*{Signed Stochastic Block Model (SSBM) and Bayesian Information Criterion (BIC)}

\begin{figure*}[ht!]
\centering
\includegraphics[width=0.9\textwidth]{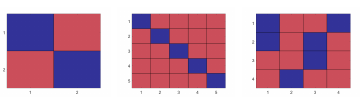}
\caption{Three, ideal partitions recoverable upon minimising BIC: one compatible with the strong balance theory ($k=2$ with $p_{11}^+>p_{11}^-$, $p_{22}^+>p_{22}^-$ and $p_{12}^+<p_{12}^-$ - left panel), one compatible with the weak balance theory ($k>2$ with $p_{rr}^+>p_{rr}^-$, $r=1\dots5$, $p_{rs}^+<p_{rs}^-$, $r,s=1\dots5$, $\forall\:r<s$ - middle panel), one compatible with the relaxed balance theory ($p_{rr}^+\leq p_{rr}^-$ for some, diagonal blocks and $p_{rs}^+\geq p_{rs}^-$ for some, off-diagonal blocks - right panel).}
\label{fig_A_1}
\end{figure*}

Let us, first, recall the derivation of the SSBM. It is defined by the Hamiltonian

\begin{align}
H(\mathbf{A})&=\sum_{r\leq s}[\alpha_{rs} L_{rs}^+(\mathbf{A})+\beta_{rs} L_{rs}^-(\mathbf{A})]\nonumber\\
&=\sum_{r\leq s}\left\{\alpha_{rs}\left[\sum_{i=1}^N\sum_{j(>i)}\delta_{g_ir}\delta_{g_js}a_{ij}^+\right]+\beta_{rs}\left[\sum_{i=1}^N\sum_{j(>i)}\delta_{g_ir}\delta_{g_js}a_{ij}^-\right]\right\}\nonumber\\
&=\sum_{i=1}^N\sum_{j(>i)}[\alpha_{g_ig_j} a_{ij}^++\beta_{g_ig_j} a_{ij}^-]
\end{align}
leading to

\begin{align}
Z=\sum_{\mathbf{A}\in\mathbb{A}}e^{-H(\mathbf{A})}=\sum_{\mathbf{A}\in\mathbb{A}}\prod_{i=1}^N\prod_{j(>i)}e^{-[\alpha_{g_ig_j} a_{ij}^++\beta_{g_ig_j} a_{ij}^-]}&=\prod_{i=1}^N\prod_{j(>i)}\sum_{a_{ij}=-1,0,+1}e^{-[\alpha_{g_ig_j} a_{ij}^++\beta_{g_ig_j} a_{ij}^-]}\nonumber\\
&=\prod_{i=1}^N\prod_{j(>i)}[1+e^{-\alpha_{g_ig_j}}+e^{-\beta_{g_ig_j}}].
\end{align}

As a consequence, 

\begin{align}
P_\text{SSBM}(\mathbf{A})=\frac{e^{-H(\mathbf{A})}}{Z}=\frac{\prod_{i=1}^N\prod_{j(>i)}e^{-[\alpha_{g_ig_j} a_{ij}^++\beta_{g_ig_j}]}}{\prod_{i=1}^N\prod_{j(>i)}[1+e^{-\alpha_{g_ig_j}}+e^{-\beta_{g_ig_j}}]}&\equiv\prod_{i=1}^N\prod_{j(>i)}\frac{x_{g_ig_j}^{a_{ij}^+}y_{g_ig_j}^{a_{ij}^-}}{1+x_{g_ig_j}+y_{g_ig_j}}\nonumber\\
&\equiv\prod_{i=1}^N\prod_{j(>i)}(p_{g_ig_j}^+)^{a_{ij}^+}(p_{g_ig_j}^0)^{a_{ij}^0}(p_{g_ig_j}^-)^{a_{ij}^-}
\end{align}
having posed $e^{-\alpha_{g_ig_j}}\equiv x_{g_ig_j}$, $e^{-\beta_{g_ig_j}}\equiv y_{g_ig_j}$, $p_{ij}^+\equiv x_{g_ig_j}/(1+x_{g_ig_j}+y_{g_ig_j})$, $p_{ij}^-\equiv y_{g_ig_j}/(1+x_{g_ig_j}+y_{g_ig_j})$, $p_{ij}^0\equiv 1/(1+x_{g_ig_j}+y_{g_ig_j})$; let us notice that

\begin{align}
P_\text{SSBM}(\mathbf{A})&=\prod_{i=1}^N\prod_{j(>i)}\prod_{r=1}^k\prod_{s(\geq r)}[(p_{rs}^+)^{a_{ij}^+}(p_{rs}^0)^{a_{ij}^0}(p_{rs}^-)^{a_{ij}^-}]^{\delta_{g_ir}\delta_{g_js}}\nonumber\\
&=\prod_{r=1}^k\prod_{s(\geq r)}\prod_{i=1}^N\prod_{j(>i)}[(p_{rs}^+)^{\delta_{g_ir}\delta_{g_js}a_{ij}^+}(p_{rs}^0)^{\delta_{g_ir}\delta_{g_js}a_{ij}^0}(p_{rs}^-)^{\delta_{g_ir}\delta_{g_js}a_{ij}^-}]\nonumber\\
&=\prod_{r=1}^k\prod_{s(\geq r)}[(p_{rs}^+)^{\sum_{i=1}^N\sum_{j(>i)}\delta_{g_ir}\delta_{g_js}a_{ij}^+}(p_{rs}^0)^{\sum_{i=1}^N\sum_{j(>i)}\delta_{g_ir}\delta_{g_js}a_{ij}^0}(p_{rs}^-)^{\sum_{i=1}^N\sum_{j(>i)}\delta_{g_ir}\delta_{g_js}a_{ij}^-}]\nonumber\nonumber\\
&=\prod_{r=1}^k(p_{rr}^+)^{L_{rr}^+}(p_{rr}^0)^{L_{rr}^0}(p_{rr}^-)^{L_{rr}^-}\cdot\prod_{r=1}^k\prod_{s(>r)}(p_{rs}^+)^{L_{rs}^+}(p_{rs}^0)^{L_{rs}^0}(p_{rs}^-)^{L_{rs}^-}\nonumber\\
&=\prod_{r=1}^k(p_{rr}^+)^{L_{rr}^+}(p_{rr}^-)^{L_{rr}^-}(1-p_{rr}^+-p_{rr}^-)^{\binom{N_r}{2}-L_{rr}}\cdot\prod_{r=1}^k\prod_{s(>r)}(p_{rs}^+)^{L^+_{rs}}(p_{rs}^-)^{L^-_{rs}}(1-p_{rs}^+-p_{rs}^-)^{N_rN_s-L_{rs}}\nonumber\\
&=\prod_{r=1}^k\frac{x_{rr}^{L_{rr}^+}y_{rr}^{L_{rr}^-}}{[1+x_{rr}+y_{rr}]^{\binom{N_r}{2}}}\cdot\prod_{r=1}^k\prod_{s(>r)}\frac{x_{rs}^{L_{rs}^+}y_{rs}^{L_{rs}^-}}{[1+x_{rs}+y_{rs}]^{N_r N_s}}
\end{align}
where $L_{rr}=L_{rr}^++L_{rr}^-$, $L_{rs}=L_{rs}^++L_{rs}^-$, $p_{rr}^+=x_{rr}/(1+x_{rr}+y_{rr})$, $p_{rr}^-=y_{rr}/(1+x_{rr}+y_{rr})$, $r=1\dots k$ and $p_{rs}^+=x_{rs}/(1+x_{rs}+y_{rs})$, $p_{rs}^-=y_{rs}/(1+x_{rs}+y_{rs})$, $\forall\:r<s$. Its log-likelihood reads

\begin{figure*}[t!]
\centering
\includegraphics[width=\textwidth]{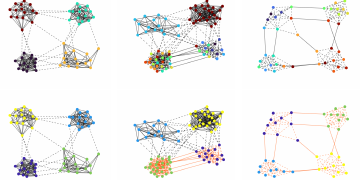}
\caption{Top panels: partitions recovered upon minimising $F(\bm{\sigma})$. Bottom panels: partitions recovered upon minimising BIC (orange links would be classified as misplaced according to the $F$-test). Minimising BIC leads us to find a community structure whose definition depends on the `signed density' of connections: such a structure coincides with the one recovered upon minimising $F(\bm{\sigma})$ only if the former is $k$-balanced, i.e. satisfies the relationships $p_{rr}^-=0$, $r=1\dots k$ and $p_{rs}^+=0$, $\forall\:r<s$.}
\label{fig_A_2}
\end{figure*}

\begin{align}
\ln\mathcal{L}_\text{SSBM}=&\sum_{r=1}^k\left[L_{rr}^+(\mathbf{A})\ln x_{rr}+L_{rr}^-(\mathbf{A})\ln y_{rr}-\binom{N_r}{2}\ln[1+x_{rr}+y_{rr}]\right]\nonumber\\
&+\sum_{r=1}^k\sum_{s(>r)}[L_{rs}^+(\mathbf{A})\ln x_{rs}+L_{rs}^-(\mathbf{A})\ln y_{rs}-N_rN_r\ln[1+x_{rs}+y_{rs}]]
\end{align}
and its maximisation leads to recover the conditions $p_{rr}^+=2L_{rs}^+(\mathbf{A})/N_r(N_r-1)$, $p_{rr}^-=2L_{rs}^-(\mathbf{A})/N_r(N_r-1)$, $r=1\dots k$ and $p_{rs}^+=L_{rs}^+(\mathbf{A})/N_rN_s$, $p_{rs}^-=L_{rs}^-(\mathbf{A})/N_rN_s$, $\forall\:r<s$.\\

Let us, now, recall that BIC is defined as

\begin{equation}
\text{BIC}=\kappa\ln n-2\ln\mathcal{L}
\end{equation}
where $\mathcal{L}$ is a model likelihood and $\kappa$ indicates the number of parameters entering into its definition. Here, we have posed $\mathcal{L}\equiv\mathcal{L}_\text{SSBM}$, $\kappa\equiv\kappa_\text{SSBM}=k(k+1)$ (i.e. $k+k(k-1)/2$ parameters to be tuned on the set of values $L_{rs}^+$, $\forall\:r\leq s$ and $k+k(k-1)/2$ parameters to be tuned on the set of values $L_{rs}^-$, $\forall\:r\leq s$) and $n=N(N-1)/2$.

Figure \ref{fig_A_1} shows three, ideal partitions: one compatible with the strong balance theory, one compatible with the weak balance theory and one compatible with the relaxed balanced theory.

\subsection*{Comparing BIC minimisation with $F$ minimisation}

Let us, now, carry out another comparison between the recipe prescribing to minimise $F(\bm{\sigma})$ and the one prescribing to minimise BIC. As fig. \ref{fig_A_2} shows. the partitions that are recovered upon minimising BIC match the planted ones, a result confirming that BIC is sensitive to the `signed density' of the modules. As a consequence, the partitions recovered upon minimising it coincide with the ones recovered upon minimising $F(\bm{\sigma})$ only if the former ones are $k$-balanced, i.e. satisfy the relationships $p_{rr}^-=0$, $r=1\dots k$ and $p_{rs}^+=0$, $\forall\:r<s$.

\begin{figure*}[t!]
\centering
\includegraphics[width=\textwidth]{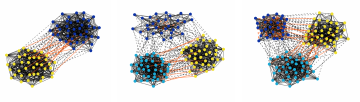}
\caption{Minimising $F(\bm{\sigma})$ can lead to a number of ambiguous situations, such as returning configurations that are neither traditionally nor relaxedly balanced (orange links are classified as misplaced according to the $F$-test; however, they can neither be arranged into homogeneous blocks). Within our, novel, statistical framework these configurations are unambiguously classified as balanced according to the statistical variant of the traditional balance theory.}
\label{fig_A_3}
\end{figure*}

Minimising $F(\bm{\sigma})$ can lead to a number of ambiguous situations, such as \emph{i)} returning configurations that are neither traditionally nor relaxedly balanced; \emph{ii)} returning more than one frustrated configuration.

Figure \ref{fig_A_3} depicts the first situation. Nodes of the same colour are those put together as a consequence of minimising $F(\bm{\sigma})$: although the presence of negative links between and within modules makes the recovered partitions `traditionally' frustrated, the original formulation of the relaxed balance theory would lead us to conclude that they are `relaxedly' frustrated as well; only within our, novel, statistical framework these configurations can be unambiguously classified as balanced according to the statistical variant of the traditional balance theory.

\begin{figure*}[t!]
\centering
\includegraphics[width=0.9\textwidth]{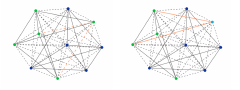}
\caption{Minimising $F(\bm{\sigma})$ can lead to a number of ambiguous situations, such as returning more than one configuration in correspondence of which $F(\bm{\sigma})$ attains its minimum. In this, particular case, the Slovenian Parliament admits two, different arrangements of nodes characterised by $F(\bm{\sigma})=2$ (orange links are classified as misplaced according to the $F$-test). Our BIC-based test, however, `prefers' the one on the left, classifying it as balanced according to the statistical variant of the traditional balance theory.}
\label{fig_A_4}
\end{figure*}

Figure \ref{fig_A_4} depicts the second situation: the minimisation of $F(\bm{\sigma})$ can return more than one frustrated configuration; our BIC-based test, however, `prefers' the one on the left, classifying it as balanced according to the statistical variant of the traditional balance theory.

\subsection*{Minimising BIC on configurations generated by the Signed Random Graph Model (SRGM)}

Let us, first, recall the derivation of the SRGM. It is defined by the Hamiltonian

\begin{equation}
H(\mathbf{A})=\alpha L^+(\mathbf{A})+\beta L^-(\mathbf{A})=\sum_{i=1}^N\sum_{j(>i)}[\alpha a_{ij}^++\beta a_{ij}^-]
\end{equation}
leading to

\begin{align}
Z=\sum_{\mathbf{A}\in\mathbb{A}}e^{-H(\mathbf{A})}=\sum_{\mathbf{A}\in\mathbb{A}}\prod_{i=1}^N\prod_{j(>i)}e^{-[\alpha a_{ij}^++\beta a_{ij}^-]}&=\prod_{i=1}^N\prod_{j(>i)}\sum_{a_{ij}=-1,0,+1}e^{-[\alpha a_{ij}^++\beta a_{ij}^-]}\nonumber\\
&=\prod_{i=1}^N\prod_{j(>i)}[1+e^{-\alpha}+e^{-\beta}]=[1+e^{-\alpha}+e^{-\beta}]^{\binom{N}{2}}.
\end{align}

As a consequence, 

\begin{equation}
P_\text{SRGM}(\mathbf{A})=\frac{e^{-H(\mathbf{A})}}{Z}=\frac{e^{-[\alpha L^+(\mathbf{A})+\beta L^-(\mathbf{A}]}}{\text{
$[1+e^{-\alpha}+e^{-\beta}]^{\binom{N}{2}}$}}\equiv\frac{x^{L^+(\mathbf{A})}y^{L^-(\mathbf{A})}}{[1+x+y]^{\binom{N}{2}}}\equiv(p^-)^{L^-(\mathbf{A})}(p^0)^{L^0(\mathbf{A})}(p^+)^{L^+(\mathbf{A})}
\end{equation}
where $p^+=x/(1+x+y)$, $p^-=y/(1+x+y)$ and $p^0=1/(1+x+y)$. Its log-likelihood reads

\begin{equation}
\ln\mathcal{L}_\text{SRGM}=L^+(\mathbf{A})\ln x+L^-(\mathbf{A})\ln y-\binom{N}{2}\ln[1+x+y]
\end{equation}
and its maximisation leads to recover the conditions $p^+=2L^+(\mathbf{A})/N(N-1)$, $p^-=2L^-(\mathbf{A})/N(N-1)$.

\begin{figure*}[t!]
\centering
\includegraphics[width=\textwidth]{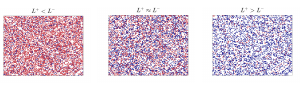}
\caption{Three configurations generated by the SRGM (positive links are coloured in blue; negative links are coloured in red). Since this model does not carry any information about a network modular structure, no groups of nodes should be detected: minimising BIC, in fact, always returns a single community gathering all nodes together, irrespectively from the choice of the parameters.}
\label{fig_A_5}
\end{figure*}

Figure \ref{fig_A_5} shows three configurations generated by the SRGM. As such a model does not carry any information about a network modular structure, no groups of nodes should be recognised. This is precisely the output of our BIC-based recipe, returning a single community gathering all nodes together (i.e. $k=1$), irrespectively from the choice of the parameters (i.e. be $p^+<p^-$, $p^+\simeq p^-$ or $p^+>p^-$). Minimising $F(\bm{\sigma})$ (or maximising $Q(\bm{\sigma})$ - see Appendix \hyperlink{AppC}{C}), instead, leads to recover a number of modules $k\geq1$ (in the case $L^+<L^-$, $k_\text{F}=25$ and $k_\text{Q}=13$; in the case $L^+\simeq L^-$, $k_\text{F}=6$ and $k_\text{Q}=5$; in the case $L^+>L^-$, $k_\text{F}=k_\text{Q}=1$).

\subsection*{Minimising BIC on negative and positive cliques}

Let us, now, consider a negative clique composed by $N$ nodes; evaluating BIC on the partition defined by $k$ modules returns the value $k(k+1)\ln[N(N-1)/2]$; since $N\geq k\geq 1$, keeping all nodes together is the most convenient choice. The same result holds true if we consider a positive clique composed by $N$ nodes, the reason lying in the completely symmetric role played by negative and positive links, both contributing to the density of the (potential) network modules.

\subsection*{Minimising BIC on complete, signed graphs}

When dealing with complete graphs, the signs come into play in a quite peculiar fashion. Let us, in fact, consider a complete graph of size $N=N_1+N_2$, constituted by two, negative cliques having, respectively, $N_1$ and $N_2$ nodes and such that each node of a clique is connected to each node of the other via a positive link. Let us, now, pose $k>2$ and consider the following inequality

\begin{equation}
k(k+1)\ln\binom{N}{2}-2\ln\mathcal{L}_\text{SSBM}>6\ln\binom{N}{2}
\end{equation}
stating that evaluating BIC on a generic partition defined by $k>2$ modules returns a value that is strictly larger than the value of BIC calculated on the bi-partition whose modules coincide with the cliques themselves (in fact, $N\geq k>2$ and $\ln\mathcal{L}_\text{SSBM}>0$).

\begin{figure*}[t!]
\centering
\includegraphics[width=0.75\textwidth]{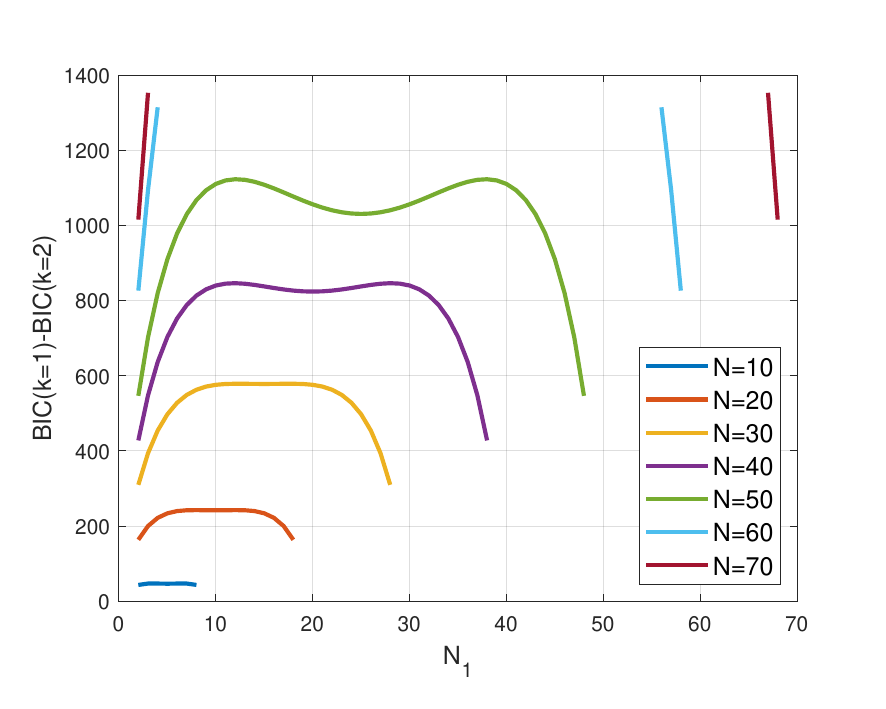}
\caption{Plotting the difference between $\text{BIC}(k=1)$ and the value of BIC characterised by the bi-partition whose modules coincide with the cliques themselves, as a function of $2<N_1<N-2$, reveals it to be always positive. This result confirms that such a bi-partition is the one in correspondence of which BIC attains its minimum value.}
\label{fig_A_6}
\end{figure*}

Let us, now, compare the bi-partition induced by the negative cliques with the partition induced by imposing the presence of just one module: in this, last case, evaluating BIC returns the value

\begin{align}
\text{BIC}(k=1)&=2\ln\binom{N}{2}-2\ln\left[(p^+)^{L^+}(p^-)^{L^-}\right]\nonumber\\
&=2\ln\binom{N}{2}-2\ln\left[\left(\frac{2N_1N_2}{N(N-1)}\right)^{N_1N_2}\left(\frac{N_1(N_1-1)+N_2(N_2-1)}{N(N-1)}\right)^{\binom{N_1}{2}+\binom{N_2}{2}}\right]\nonumber\\
&=2\ln\binom{N}{2}-2\ln\left[\left(\frac{2N_1(N-N_1)}{N(N-1)}\right)^{N_1(N-N_1)}\left(\frac{N_1(N_1-1)+(N-N_1)(N-N_1-1)}{N(N-1)}\right)^{\binom{N_1}{2}+\binom{N-N_1}{2}}\right]
\end{align}
where we have used the relationship $N_2=N-N_1$. As depicted in fig. \ref{fig_A_6}, splitting nodes according to the partition induced by the signs is always `more convenient' than partitioning them in a different way, a result suggesting that signs keep playing a role as long as the information embodied by the network density becomes irrelevant.

\begin{figure*}[b!]
\centering
\includegraphics[width=\textwidth]{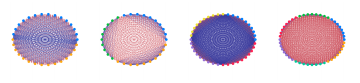}
\caption{Results of testing BIC minimisation on four, complete, signed graphs balanced according to the traditional balance theory (second panel: a 3-balanced configuration; fourth panel: a 6-balanced configuration) and maximally frustrated according to the traditional balance theory but perfectly balanced according to the relaxed balance theory (first panel: $k=2$; third panel: $k=4$). Minimising BIC leads to the partition induced by signs, since always `more convenient' than any, other partition. Positive links are coloured in blue; negative links are coloured in red.}
\label{fig_A_7}
\end{figure*}

Purely numerical experiments seem to suggest that an analogous result holds true for any number of cliques,
% \begin{align}
% c(c+1)\ln\binom{N}{2}&<2\ln\binom{N}{2}-2\ln\left[\left(\frac{2\sum_{i<j}N_iN_j}{N(N-1)}\right)^{\sum_{i<j}N_iN_j}\left(\frac{\sum_i N_i(N_i-1)}{N(N-1)}\right)^{\sum_i\binom{N_i}{2}}\right],
% \end{align}
i.e. the partition defined by (the modules that coincide with) the cliques themselves is the one attaining the minimum value of BIC (see also fig. \ref{fig_A_7}).

\clearpage

\begin{figure*}[t!]
\centering
\includegraphics[width=\textwidth]{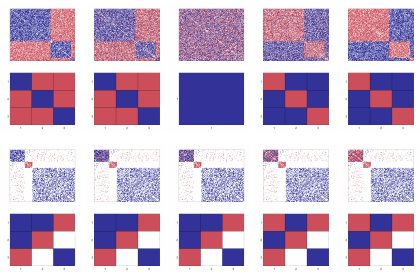}
\caption{Partitions recovered upon minimising BIC on, five, synthetic (dense: upper panels; sparse: lower panels) configurations generated via complementary probability matrices. By varying a single parameter, the $C=3$ blocks can be designed to obey either the traditional or the relaxed balance theory. Upper panels: the sizes of the blocks are $100$, $50$ and $10$ while the parameters have been set to the values $c=0.7$ and $\alpha=0.7,0.6,0.35,0.1,0$. Lower panels: the sizes of the blocks are $30$, $15$ and $85$ while the parameters have been set to the values $c=0.8$ and $\alpha=0.8,0.6,0.4,0.2,0$.}
\label{fig_A_8}
\end{figure*}

\subsection*{Minimising BIC on more, synthetic configurations}

More realistic, planted partitions can be obtained as follows. Let us consider a matrix with $C$ communities, inducing a $C\times C$ block matrices. Let us, now, pose $p^+_\bullet=\alpha$, $p^-_\bullet=c-\alpha$, $p^+_\circ=c-\alpha$, $p^-_\circ=\alpha$ with $\alpha$ playing the role of internal density of positive links and $c$ playing the role of link density: more formally,

\begin{align}
\mathbf{P^+}=
\begin{bmatrix}
\alpha & c-\alpha & c-\alpha \\
c-\alpha & \alpha & c-\alpha \\
c-\alpha & c-\alpha & \alpha \\
\end{bmatrix},
\quad
\mathbf{P^-}=
\begin{bmatrix}
c-\alpha & \alpha & \alpha \\
\alpha & c-\alpha & \alpha \\
\alpha & \alpha & c-\alpha \\
\end{bmatrix}.
\end{align} 

Considering probability matrices that are complementary allows us to span a wide spectrum of different configurations, by perturbing the entire network structure at the same time (from this perspective, the considered benchmark is similar-in-spirit to the so-called Aldecoa's `relaxed caveman' benchmark): by fixing $c$ and letting $\alpha$ range from $0$ to $c$, in fact, one can move from traditionally balanced networks to relaxedly balanced networks, through configurations that are compatible with the SRGM. The upper panels of fig.~\ref{fig_A_8} have been produced by considering $C=3$ blocks of size $100$, $50$, $10$ and setting $c=0.7$, $\alpha=0.7,0.6,0.35,0.1,0$.

The lower panels of fig.~\ref{fig_A_8} have been produced by considering the setting

\begin{align}
\mathbf{P^+}=
\begin{bmatrix}
\alpha & 0.05 & 0 \\
0.05 & 0 & 0 \\
0 & 0 & 0.4 \\
\end{bmatrix},
\quad
\mathbf{P^-}=
\begin{bmatrix}
c-\alpha & 0 & 0.05 \\
0 & 0.8 & 0 \\
0.05 & 0 & 0 \\
\end{bmatrix},
\end{align}
identifying $C=3$ blocks whose sizes are $30$, $15$ and $85$, with $c=0.8$ and where the only varying parameter is $\alpha\in\{0.8,0.6,0.4,0.2,0\}$: the partitions recovered by minimising BIC are always consistent with the planted ones, irrespectively from the density of (the blocks constituting) the considered configuration.

\begin{figure*}[t!]
\centering
\includegraphics[width=\textwidth]{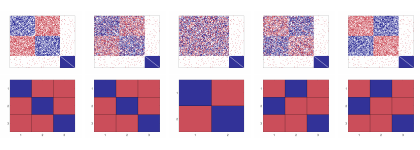}
\caption{Partitions recovered upon minimising BIC on, five, synthetic configurations generated via complementary probability matrices. By varying a single parameter, the $C=3$ blocks (whose sizes are $50$, $50$ and $30$) can be designed to obey either the traditional or the relaxed balance theory. The parameters have been set to the values $c=0.6$ and $\alpha=0.6,0.45,0.3,0.15,0$.}
\label{fig_A_9}
\end{figure*}

Finally, the panels of fig.~\ref{fig_A_9} have been produced by considering the setting

\begin{align}
\mathbf{P^+}=
\begin{bmatrix}
\alpha & c-\alpha & 0 \\
c-\alpha & \alpha & 0 \\
0 & 0 & 1 \\
\end{bmatrix},
\quad
\mathbf{P^-}=
\begin{bmatrix}
c-\alpha & \alpha & 0.03 \\
\alpha & c-\alpha & 0.03 \\
0.03 & 0.03 & 1 \\
\end{bmatrix},
\end{align}
identifying $C=3$ blocks whose sizes are $50$, $50$ and $30$, with $c=0.6$ and where the only varying parameter is $\alpha\in\{0.6,0.45,0.3,0.15,0\}$: the partitions recovered by minimising BIC are always consistent with the planted ones across a wide range of the parameters values.

\clearpage

\begin{figure*}[t!]
\centering
\includegraphics[width=\textwidth]{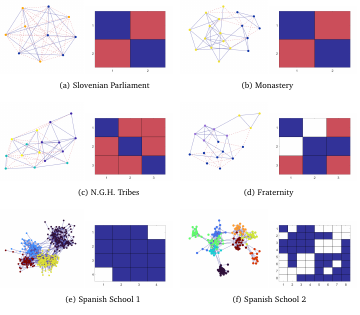}
\caption{Partitions recovered upon minimising BIC on the Slovenian Parliament~\cite{doreian1996partitioning}, Monastery, N.G.H. Tribes, Fraternity~\cite{aref2019balance} and Spanish Schools~\cite{ruiz2023triadic} datasets. A generic block, indexed as $rs$, is coloured in blue if $L_{rs}^+>L_{rs}^-$, in red if $L_{rs}^+<L_{rs}^-$ and in white if $L_{rs}^+=L_{rs}^-$. Minimising BIC leads to recover partitions that obey either the statistical variant of the traditional balance theory or the statistical variant of the relaxed balance theory. Positive links are coloured in blue; negative links are coloured in red.}
\label{fig_A_10}
\end{figure*}

\subsection*{Minimising BIC on more, real-world configurations}

Let us, now, apply our recipe to a number of real-world, signed configurations, i.e. Slovenian Parliament~\cite{doreian1996partitioning}, Monastery, N.G.H. Tribes, Fraternity~\cite{aref2019balance} and Spanish Schools~\cite{ruiz2023triadic}. Since the last dataset originally displayed directed interactions, we have made it undirected by applying the following set of rules: `$+/+$' becomes `$+$' and `$-/-$' becomes `$-$' (i.e. if any two agents have the same opinion, the undirected connection preserve the sign); `$+/-$' becomes `$-$' (i.e. if any two agents have opposite opinions, the undirected connection has a negative sign); `$+/0$' and `$-/0$' become `$0$' (i.e. any interaction that is not reciprocated disappears in the undirected version of the network).

As fig. \ref{fig_A_10} shows, minimising BIC leads to recover partitions that obey either the statistical variant of the traditional balance theory (in its strong or weak form) or the statistical variant of the relaxed balance theory (a blue block is characterised by a majority of positive links; a red block is characterised by a majority of negative links). On the other hand, US Senate, EGFR, Macrophage, Bitcoin Alpha and Bitcoin OTC datasets~\cite{aref2020multilevel} are characterised by $k=1$ (either because $p^+\simeq p^-$ or because $p^+\gg p^-$).

The community structure characterising synthetic and real-world, signed networks has been also explored in~\cite{kunegis2010spectral,anchuri2012communities,esmailian2014mesoscopic,jiang2015stochastic,yang2017stochastic,ping2019community,zhong2022efficient,zhang2023community,diaz2024signed}.\\

\begin{figure*}[t!]
\centering
\includegraphics[width=\textwidth]{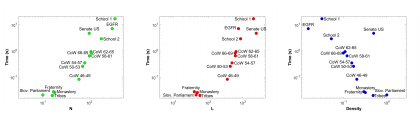}
\caption{The computational complexity of the algorithm to minimise BIC rises with $N$ and $L$ but decreases with the link density.}
\label{fig_A_11}
\end{figure*}

From a purely computational perspective, the complexity of the algorithm to minimise BIC rises with $N$ and $L$ although it decreases with the link density $c(\mathbf{A})=2L(\mathbf{A})/N(N-1)$, as fig. \ref{fig_A_11} shows - in fact, the denser the configuration, the faster the algorithm.

\clearpage

\hypertarget{AppC}{}
\section*{Appendix C.\\More on signed modularity}

A deterministic theory of balance can be turned into a statistical theory of balance by answering the following question: is it possible to define a reference level of misplaced links by means of which discerning frustrated graphs from balanced ones? The answer is affirmative and calls for comparing the empirical amount of frustration of a given, signed configuration with the one predicted by a properly-defined reference model: in formulas, one may define a partition as \emph{statistically balanced} if satisfying the relationship $F(\bm{\sigma})<\langle F(\bm{\sigma})\rangle$.

A quantity embodying such a comparison already exists: it is the \emph{signed modularity}, reading

\begin{align}
Q(\bm{\sigma})&=\sum_{i=1}^N\sum_{j(>i)}[(a_{ij}^+-p_{ij}^+)-(a_{ij}^--p_{ij}^-)]\delta_{\sigma_i,\sigma_j}\nonumber\\
&=L_\bullet^+-\langle L_\bullet^+\rangle-(L_\bullet^--\langle L_\bullet^-\rangle)\nonumber\\
&=(L^+-L_\circ^+)-\langle L^+-L_\circ^+\rangle-(L_\bullet^--\langle L_\bullet^-\rangle)\nonumber\\
&=-(L_\circ^++L_\bullet^-)+\langle L_\circ^++L_\bullet^-\rangle+L^+-\langle L^+\rangle\nonumber\\
&=-F(\bm{\sigma})+\langle F(\bm{\sigma})\rangle+L^+-\langle L^+\rangle
\end{align}
with obvious meaning of the symbols (the addendum $L^+-\langle L^+\rangle$ is just an offset not depending on the specific partition and amounting at zero for any model reproducing the total number of positive links)~\cite{gomez2009analysis}. Since the total number of positive links is preserved under any model considered in the present paper, we obtain

\begin{equation}
Q(\bm{\sigma})=-F(\bm{\sigma})+\langle F(\bm{\sigma})\rangle.
\end{equation}

$Q(\bm{\sigma})$ has been widely employed to spot communities on signed networks, with the positions $p_{ij}^+=k_i^+k_j^+/2L^+$ and $p_{ij}^-=k_i^-k_j^-/2L^-$, $\forall\:i<j$~\cite{anchuri2012communities,gomez2009analysis,amelio2013community}. Such a recipe, instantiating the Chung-Lu (CL) model, is applicable only in case $p_{ij}^+\leq1$ and $p_{ij}^-\leq1$, $\forall\:i<j$: these conditions, however, do not hold in several cases of interest, an example of paramount importance being provided by sparse networks whose degree distribution is scale-free~\cite{vallarano2021}. In order to overcome the aforementioned limitation, a different framework is needed.

One may follow the analytical approach introduced in~\cite{park2004statistical} and further developed in~\cite{Squartinia}, aimed at identifying the functional form of the maximum-entropy probability distribution that preserves a desired set of empirical constraints, on average. Specifically, this approach looks for the graph probability $P(\mathbf{A})$ that maximises Shannon entropy $S=-\sum_{\mathbf{A}\in\mathbb{A}}P(\mathbf{A})\ln P(\mathbf{A})$, under constraints enforcing the expected value of a chosen set of properties. The formal solution to this problem is the exponential probability $P(\mathbf{A})=e^{-H(\mathbf{A})}/Z$ where the \emph{Hamiltonian} $H(\mathbf{A})$ is a linear combination of the constrained properties and the \emph{partition function} $Z=\sum_{\mathbf{A}\in\mathbb{A}}e^{-H(\mathbf{A})}$ plays the role of normalising constant, the sum running over the set $\mathbb{A}$ of all binary, undirected, signed graphs whose cardinality amounts to $|\mathbb{A}|=3^{\binom{N}{2}}$. Two examples of models of the kind are the Signed Random Graph Model (SRGM) and the Signed Configuration Model (SCM)~\cite{gallo2024testing}.

According to the traditional balance theory, several ways exist in which a given configuration can be frustrated. Let us, now, analyse them in detail.

\subsubsection{Evaluating frustration due to negative links}

\paragraph*{Positive subgraphs connected by negative links.} In order to understand how a $Q$-based test would work, let us consider two subgraphs with, respectively, $m$ and $n$ nodes, positive intra-modular links and negative inter-modular links. Let us denote with $V_\bullet=m(m-1)/2+n(n-1)/2$ the total number of intra-modular pairs of nodes and with $Q_0$ the value of modularity associated to the partition of the entire graph, except our, two subgraphs; let us also call $L_\bullet^+$ the total number of positive links within modules and $L_\circ^-$ the total number of negative links between modules. Then,

\begin{align}
Q_A&=Q_0+[L_\bullet^+-p^+V_\bullet]-[0-p^-V_\bullet],\\
Q_B&=Q_0+[L_\bullet^+-p^+(V_\bullet+mn)]-[L_\circ^--p^-(V_\bullet+mn)]
\end{align}
with $Q_A$ being the SRGM-induced modularity of the configuration identifying our subgraphs as two, separate communities and $Q_B$ being the SRGM-induced modularity of the configuration identifying our subgraphs as a single community. In order to be fully consistent with the traditional balance theory, one should require

\begin{equation}
Q_A>Q_B=Q_A-L_\circ^--(p^+-p^-)mn,
\end{equation}
a condition that it is satisfied whenever $L_\circ^->(p^--p^+)mn$, i.e. whenever the probability $p_\circ^-\equiv L_\circ^-/mn$ of establishing a negative link within modules is larger than $p^--p^+=2(L^--L^+)/N(N-1)$. This condition sheds light on the role played by the signed variant of the resolution limit, naturally re-interpretable as a threshold-based criterion for discerning if a given, signed configuration is balanced or not: in words, the `acceptable' level of frustration, according to which our subgraphs can be safely interpreted as a single community, is represented by $(p^--p^+)mn$.

\subsubsection{Evaluating frustration due to positive links}

\paragraph*{Positive subgraphs connected by positive links.} Let us, now, consider two subgraphs with, respectively, $m$ and $n$ nodes, positive intra- and inter-modular links; let us also call $L_\bullet^+$ the total number of positive links within modules and $L_\circ^+$ the total number of positive links between modules. Then,

\begin{align}
Q_A&=Q_0+[L_\bullet^+-p^+V_\bullet]-[0-p^-V_\bullet],\\
Q_B&=Q_0+[L_\bullet^++L_\circ^+-p^+(V_\bullet+mn)]-[0-p^-(V_\bullet+mn)]
\end{align}
with $Q_A$ being the SRGM-induced modularity of the configuration identifying our subgraphs as two, separate communities and $Q_B$ being the SRGM-induced modularity of the configuration identifying our subgraphs as a single community. In order to be fully consistent with the traditional balance theory, one should require

\begin{equation}
Q_B=Q_A+L_\circ^+-(p^+-p^-)mn>Q_A,
\end{equation}
a condition that it is satisfied whenever $L_\circ^+>(p^+-p^-)mn$, i.e. whenever the probability $p_\circ^+\equiv L_\circ^+/mn$ of establishing a positive link between modules is larger than $p^+-p^-=2(L^+-L^-)/N(N-1)$. The threshold-based criterion for discerning balance represented by the signed variant of the resolution limit, now, sets the `acceptable' level of frustration, according to which our subgraphs can be safely interpreted as two, separate communities, at $(p^+-p^-)mn$.

\begin{figure*}[t!]
\centering
\includegraphics[width=\textwidth]{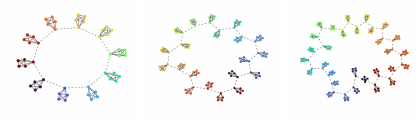}
\caption{Partitions recovered upon maximising the signed modularity on three rings of cliques, i.e. a set of 10 (left), 20 (middle), 30 (right) cliques, constituted by 5 nodes each, internally connected by positive links and inter-connected by negative links.}
\label{fig_A_12}
\end{figure*}

\subsubsection{Evaluating frustration in real-world networks}

Interestingly enough, when studying real-world, signed networks, the relationship $L^+\gg L^-$ is often (if not always) found to hold true: as a consequence, the condition

\begin{align}\label{eqneg}
L_\circ^->(p^--p^+)mn<0
\end{align}
is trivially satisfied. Such an evidence has several consequences. In order to discuss them, let us focus on \emph{i)} the case of negative subgraphs connected by negative links; \emph{ii)} the case of negative subgraphs connected by positive links.\\

\begin{figure*}[t!]
\includegraphics[width=\textwidth]{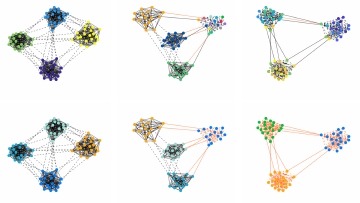}
\caption{Top panels: partitions recovered upon maximising $Q(\bm{\sigma})$. Bottom panels: partitions recovered upon minimising BIC. Left panels: minimising BIC returns partitions that coincide with those returned upon maximising the signed modularity only in case they are $k$-balanced. Middle and right panels: minimising BIC (bottom panels) leads to recover the planted partitions, balanced according to the relaxed balance theory; maximising $Q(\bm{\sigma})$ (top panels), instead, leads to the fragmentation of the subgraphs constituted by negative links into singletons. In other words, the $Q$-based test seeks to recover a configuration obeying the traditional balance theory even when there is none `by design' (orange links are classified as misplaced according to the $F$-test).}
\label{fig_A_13}
\end{figure*}

\paragraph*{Negative subgraphs connected by negative links.} Let us consider two subgraphs with, respectively, $m$ and $n$ nodes, negative intra- and inter-modular links; let us also call $L_\bullet^-$ the total number of negative links within modules and $L_\circ^-$ the total number of negative links between modules. Then,

\begin{align}
Q_A&=Q_0+[0-p^+V_\bullet]-[L_\bullet^--p^-V_\bullet],\\
Q_B&=Q_0+[0-p^+(V_\bullet+mn)]-[L_\bullet^-+L_\circ^--p^-(V_\bullet+mn)]
\end{align}
with $Q_A$ being the SRGM-induced modularity of the configuration identifying our subgraphs as two, separate communities and $Q_B$ being the SRGM-induced modularity of the configuration identifying our subgraphs as a single community. In order to limit the number of links that would be deemed as misplaced according to the traditional balance theory, one should require

\begin{equation}
Q_A>Q_B=Q_A-L_\circ^--(p^+-p^-)mn,
\end{equation}
a condition that it is satisfied whenever $L_\circ^->(p^--p^+)mn$, i.e. whenever the probability $p_\circ^-\equiv L_\circ^-/mn$ of establishing a negative link between modules is larger than $p^--p^+=2(L^--L^+)/N(N-1)<0$. Hence, it is always convenient to separate negatively connected modules and, if such a line of reasoning is repeated in a hierarchical fashion, it is always convenient to separate negatively connected nodes. Otherwise stated, one should not expect the presence of negative links within blocks since negatively connected modules will always lead to singletons: in this sense, the signed modularity is resolution limit-free.\\

\paragraph*{Negative subgraphs connected by positive links.} Let us, now, focus on the case of negative subgraphs connected by positive links and consider two subgraphs with, respectively, $m$ and $n$ nodes, negative intra-modular links and positive inter-modular links; let us also call $L_\bullet^-$ the total number of negative links within modules and $L_\circ^+$ the total number of positive links between modules. Then,

\begin{align}
Q_A&=Q_0+[0-p^+V_\bullet]-[L_\bullet^--p^-V_\bullet],\\
Q_B&=Q_0+[L_\circ^+-p^+(V_\bullet+mn)]-[L_\bullet^--p^-(V_\bullet+mn)]
\end{align}
with $Q_A$ being the SRGM-induced modularity of the configuration identifying our subgraphs as two, separate communities and $Q_B$ being the SRGM-induced modularity of the configuration identifying our subgraphs as a single community. In order to limit the number of links that would be deemed as misplaced according to the traditional balance theory, one should require

\begin{equation}
Q_B=Q_A+L_\circ^+-(p^+-p^-)mn>Q_A,
\end{equation}
a condition that it is satisfied whenever $L_\circ^+>(p^+-p^-)mn$, i.e. whenever the probability $p_\circ^+\equiv L_\circ^+/mn$ of establishing a positive link between modules is larger than $p^+-p^-=2(L^+-L^-)/N(N-1)$. Now, as a consequence of eq. \eqref{eqneg}, it is convenient to fragment negatively connected modules into singletons; hence, according to the traditional balance theory, frustration can only occur because of misplaced, positive links appearing between blocks.

Figure \ref{fig_A_12} depicts the results of the signed modularity maximisation on three rings of cliques: since the relationship $L^+\gg L^-$ holds true, one should not expect the presence of negative links within blocks (as we said, the signed modularity is resolution limit-free, in this sense). Notice that our exercise is defined in such a way that the numerical value of the generic addendum $(a_{ij}^+-p_{ij}^+)-(a_{ij}^--p_{ij}^-)$ is fixed, once and for all, by the choice of the benchmark to be solved: in other words, the definition of modularity does not change with the level of aggregation, being just recomputed (as any other score function) as the partition changes~\cite{gallo2024testing}.

As fig. \ref{fig_A_13} shows, minimising BIC returns partitions that coincide with those returned by maximising modularity, or minimising $F(\bm{\sigma})$, solely in case they are $k$-balanced (i.e. obey the traditional balance theory). In case subgraphs constituted by negative links are, instead, present, minimising BIC leads to the planted partition induced by gathering such nodes together while maximising $Q(\bm{\sigma})$ leads to their fragmentation, hence recovering singletons. In other words, the $Q$-based test (exactly as the $F$-based test and the $G$-based test) seeks to recover traditionally balanced configurations even when there is none `by design'.

From a purely numerical perspective, partitioning nodes by minimising BIC is accomplished as described in Algorithms \ref{alg:minBIC1} - \ref{alg:upBIC}.

For further details on the signed modularity, see also~\cite{gomez2009analysis,traag2009community,amelio2013community}.

\clearpage

\begin{algorithm}[ht!]
\caption{Pseudocode to partition nodes by minimising BIC - step I}
\begin{algorithmic}
\item[\hspace{1.4pt} 1:
\textbf{function} \textit{BICBasedCommunityDetectionStepI}$(N,\mathbf A)$]
\item[\hspace{1.4pt} 2: $C$ $\leftarrow$ array of labels of length $N$, initialised as $(1,2\dots N)$;]
\item[\hspace{1.4pt} 3: $\text{BIC} \leftarrow$ \textit{UpdateBIC}$(\mathbf A,C)$;]
\item[\hspace{1.4pt} 4: $E \leftarrow$ randomly sorted edges;] 
\item[\hspace{1.4pt} 5: \textbf{for} $(u,v)\in E$ \textbf{do}]
\item[\hspace{1.4pt} 6: \hspace{15pt} $C_0 \leftarrow C$;]
\item[\hspace{1.4pt} 7: \hspace{15pt} $\text{BIC}_0 \leftarrow \text{BIC}$;]
\item[\hspace{1.4pt} 8: \hspace{15pt} \textbf{if} $C(u)\neq C(v)$ \textbf{then}]
\item[\hspace{1.4pt} 9:  \hspace{34pt} $C_1 \leftarrow C$;]
\item[\hspace{1.4pt} 10: \hspace{30pt} \textbf{for} node $w\in C(u)$ \textbf{do}]
\item[\hspace{1.4pt} 11: \hspace{45pt} $C_1(w) \leftarrow C(v)$;]
\item[\hspace{1.4pt} 12: \hspace{30pt} \textbf{end for}]
\item[\hspace{1.4pt} 13: \hspace{30pt} $\text{BIC}_1 \leftarrow$ \textit{UpdateBIC}$(\mathbf A,C_1)$;]
\item[\hspace{1.4pt} 14: \hspace{10pt} \textbf{end if}]
\item[\hspace{1.4pt} 15: \hspace{10pt} \textbf{if} $\text{BIC}_1<\text{BIC}_0$ \textbf{then}]
\item[\hspace{1.4pt} 16: \hspace{30pt} $C \leftarrow C_1$;]
\item[\hspace{1.4pt} 17: \hspace{30pt} $\text{BIC} \leftarrow \text{BIC}_1$]
\item[\hspace{1.4pt} 18: \hspace{10pt} \textbf{else}]
\item[\hspace{1.4pt} 19: \hspace{30pt} $C \leftarrow C_0$;]
\item[\hspace{1.4pt} 20: \hspace{30pt} $\text{BIC} \leftarrow \text{BIC}_0$]
\item[\hspace{1.4pt} 21: \hspace{10pt} \textbf{end if}]
\item[\hspace{1.4pt} 22: \textbf{end for}]
\item[\hspace{1.4pt} 23: $\Rightarrow$ repeat the for-loop to improve the chance of finding the best partition]
\end{algorithmic} 
\label{alg:minBIC1}
\end{algorithm}

\begin{algorithm}[ht!]
\caption{Pseudocode to partition nodes by minimising BIC - step II}
\begin{algorithmic}
\item[\hspace{1.4pt} 1:
\textbf{function} \textit{BICBasedCommunityDetectionStepII}$(N,\mathbf A)$]
\item[\hspace{1.4pt} 2: $C$ $\leftarrow$ \textit{BICBasedCommunityDetectionStepI$(N,\mathbf A)$};]
\item[\hspace{1.4pt} 3: $\text{BIC} \leftarrow$ \textit{UpdateBIC}$(\mathbf A,C)$;]
\item[\hspace{1.4pt} 4: $E \leftarrow$ randomly sorted edges;] 
\item[\hspace{1.4pt} 5: \textbf{for} $(u,v)\in E$ \textbf{do}]
\item[\hspace{1.4pt} 6: \hspace{15pt} $C_0 \leftarrow C$; ]
\item[\hspace{1.4pt} 7: \hspace{15pt} $\text{BIC}_0 \leftarrow \text{BIC}$; ]
\item[\hspace{1.4pt} 8: \hspace{15pt} \textbf{if} $C(u)\neq C(v)$ \textbf{then}]
\item[\hspace{1.4pt} 9: \hspace{34pt} $C_1 \leftarrow C$;]
\item[\hspace{1.4pt} 10: \hspace{30pt} $C_1(u) \leftarrow C(v)$; ]
\item[\hspace{1.4pt} 11: \hspace{30pt} $\text{BIC}_1 \leftarrow$ \textit{UpdateBIC}$(\mathbf A,C_1)$;]
\item[\hspace{1.4pt} 12: \hspace{30pt} $C_2 \leftarrow C$; ]
\item[\hspace{1.4pt} 13: \hspace{30pt} $C_2(v) \leftarrow C(u)$; ]
\item[\hspace{1.4pt} 14: \hspace{30pt} $\text{BIC}_2 \leftarrow$ \textit{UpdateBIC}$(\mathbf A,C_2)$;]
\item[\hspace{1.4pt} 15: \hspace{10pt}
\textbf{else if} $C(u)=C(v)$
\textbf{then}]
\item[\hspace{1.4pt} 16: \hspace{30pt} $C_1 \leftarrow C$;]
\item[\hspace{1.4pt} 17: \hspace{30pt} $C_1(u) \leftarrow$ randomly sorted community different from $C(v)$;]
\item[\hspace{1.4pt} 18: \hspace{30pt} $\text{BIC}_1 \leftarrow$ \textit{UpdateBIC}$(\mathbf A,C_1)$;]
\item[\hspace{1.4pt} 19: \hspace{30pt} $C_2 \leftarrow C$;]
\item[\hspace{1.4pt} 20: \hspace{30pt} $C_2(v) \leftarrow$ randomly sorted community different from $C(u)$;]
\item[\hspace{1.4pt} 21: \hspace{30pt} $\text{BIC}_2 \leftarrow$ \textit{UpdateBIC}$(\mathbf A,C_2)$;]
\item[\hspace{1.4pt} 22: \hspace{10pt} \textbf{end if}]
\item[\hspace{1.4pt} 23: \hspace{10pt} $i \leftarrow \text{argmin}\{\text{BIC}_0,\text{BIC}_1,\text{BIC}_2\}$; ]
\item[\hspace{1.4pt} 24: \hspace{10pt} $C \leftarrow C_i$;]
\item[\hspace{1.4pt} 25: \hspace{10pt} $\text{BIC} \leftarrow \text{BIC}_i$;]
\item[\hspace{1.4pt} 26: \textbf{end for}]
\item[\hspace{1.4pt} 27: $\Rightarrow$ repeat the for-loop to improve the chance of finding the best partition]
\end{algorithmic} 
\label{alg:minBIC2}
\end{algorithm}

\begin{algorithm}[ht!]
\caption{Pseudocode to update BIC}
\begin{algorithmic}
\item[\hspace{1.4pt} 1: \textbf{function} \textit{UpdateBIC}$(\mathbf A,C)$]
\item[\hspace{1.4pt} 2: $k\leftarrow$ number of modules, i.e. number of distinct labels in $C$;]
\item[\hspace{1.4pt} 3: $\mathbf{P}^-\leftarrow$ $k\times k$ matrix whose entry $(c_1,c_2)$ is the probability that a node $u\in C(u)=c_1$ is linked]
\item[\hspace{42pt} via a $-1$ to a node $v\in C(v)=c_2$;]
\item[\hspace{1.4pt} 4: $\mathbf{P}^+\leftarrow$ $k\times k$ matrix whose entry $(c_1,c_2)$ is the probability that a node $u\in C(u)=c_1$ is linked]
\item[\hspace{42pt} via a $+1$ to a node $v\in C(v)=c_2$;]
\item[\hspace{1.4pt} 5: $\mathbf{L}^-\leftarrow$ $k\times k$ matrix whose entry $(c_1,c_2)$ is \emph{i)} the number of $-1$s between $c_1$ and $c_2$, if $c_1\neq c_2$;]
\item[\hspace{42pt} \emph{ii)} the number of $-1$s within $c_1$, otherwise;]
\item[\hspace{1.4pt} 6: $\mathbf{L}^+\leftarrow$ $k\times k$ matrix whose entry $(c_1,c_2)$ is \emph{i)} the number of $+1$s between $c_1$ and $c_2$, if $c_1\neq c_2$;]
\item[\hspace{42pt} \emph{ii)} the number of $+1$s within $c_1$, otherwise;]
\item[\hspace{1.4pt} 7: $\mathbf{n}\leftarrow$ $k\times 1$ array whose $c$-th entry is the number of nodes belonging to $c$;]
\item[\hspace{1.4pt} 8: $\mathcal L\leftarrow 1$;]
\item[\hspace{1.4pt} 9: \textbf{for} $c=1\dots k$ \textbf{do}]
\item[\hspace{1.4pt} 10: \hspace{15pt} $\mathcal L = \mathcal L\cdot \mathbf{P}^-(c,c)^{\mathbf{L}^-(c,c)}\mathbf{P}^+(c,c)^{\mathbf{L}^+(c,c)}(1-\mathbf{P}^-(c,c)-\mathbf{P}^+(c,c))^{\binom{\mathbf{n}(c)}{2}-\mathbf{L}^-(c,c)-\mathbf{L}^+(c,c)}$;]
\item[\hspace{1.4pt} 11: \hspace{15pt} \textbf{for} $d=c+1\dots k$ \textbf{do}]
\item[\hspace{1.4pt} 12: \hspace{30pt} $\mathcal L = \mathcal L\cdot \mathbf{P}^-(c,d)^{\mathbf{L}^-(c,d)}\mathbf{P}^+(c,d)^{\mathbf{L}^+(c,d)}(1-\mathbf{P}^-(c,d)-\mathbf{P}^+(c,d))^{\mathbf{n}(c)\mathbf{n}(d)-\mathbf{L}^-(c,d)-\mathbf{L}^+(c,d)}$;]
\item[\hspace{1.4pt} 13: \hspace{15pt} \textbf{end for}]
\item[\hspace{1.4pt} 14: \textbf{end for}]
\item[\hspace{1.4pt} 15: \text{BIC} = $k(k+1)\ln\binom{N}{2}-2\ln\mathcal L$]
\end{algorithmic} 
\label{alg:upBIC}
\end{algorithm}

\end{document}